\documentclass[a4paper,11pt]{article}
\pdfoutput=1 
\usepackage{jcappub}
\usepackage{float}
\usepackage{color}
\usepackage{xcolor}
\usepackage{booktabs}
\usepackage{amsmath}
\usepackage{wrapfig}

\usepackage{graphicx}
\usepackage{bm}% bold math
\usepackage{epstopdf}

\usepackage{subfigure}
\bibpunct{(}{)}{;}{a}{}{,}
\usepackage{subfigure}
\usepackage{amsmath}
\usepackage{amsfonts}
\usepackage{amssymb}
\usepackage{ulem} 
\usepackage{graphicx}
\usepackage{natbib}

\title{On the hadronic origin of the TeV radiation from GRB\,190114C}
\author[1,2,3]{Gagliardini; Silvia}
\author[1,2,]{Celli; Silvia}
\author[3]{Guetta; Dafne}
\author[1,2,4]{Zegarelli; Angela}
\author[2,3]{Capone; Antonio}
\author[1,2]{Di Palma; Irene}

\affiliation[1]{Dipartimento di Fisica, Universit\`a Sapienza, P. le Aldo Moro 2, Rome, Italy}
\affiliation[2]{Istituto Nazionale di Fisica Nucleare, Sezione di Roma, P. le Aldo Moro 2, Rome, Italy}
\affiliation[3]{Department of Physics, Ariel University, Ramat HaGolan St 65, Ariel, Israel}
\affiliation[4]{Dipartimento di Fisica, Università di Roma Tor Vergata, Via della Ricerca Scientifica, Rome, Italy}
\emailAdd{silvia.gagliardini@roma1.infn.it}

\abstract{
The recently discovered TeV emission from Gamma-Ray Bursts (GRBs) has renewed the long-standing discussion about the hadronic versus leptonic origin of the observed GRB radiation. In this work, we investigate the possibility that the very high energy gamma rays observed by MAGIC from GRB\,190114C (with energy from $\sim$0.1 to $\sim$0.8 TeV) are originated in a hadronic model.
We developed a Monte Carlo (MC) simulation of the source internal state dynamics and of the photo-hadronic interactions at internal shock. We included in the simulation also the pair production process that the secondary gamma rays undergo in the GRB jet. 
We find upper limits on the internal shock model parameters by comparing our simulations to the  sub-TeV observations of GRB\,190114C.
Neutrino flux predictions by the model are found to be consistent with experimental upper limits set by ANTARES and IceCube.}


\notoc

\begin{document}
\maketitle
\flushbottom
\newpage

\newcommand\T{\rule{0pt}{2.6ex}}       % Top strut
\newcommand\B{\rule[-1.2ex]{0pt}{0pt}} % Bottom strut

\def\apj{ApJ}                 % Astrophysical Journal
\def\apjls{ApJLS}               % Astrophysical Journal, Letters
\def\apjl{ApJL}               % Astrophysical Journal, Letters
\def\apjs{ApJS}               % Astrophysical Journal, Supplement
\def\cqg{CQG} 
\def\mnras{MNRAS}             % Monthly Notices of the RAS
\def\aap{A\&A}                % Astronomy and Astrophysics
\def\aaps{A\&AS}              % Astronomy and Astrophysics, Supplement
\def\aj{AJ}                   % Astronomical Journal
\def\physrep{Phys.~Rep.}      % Physics Reports
\def\nat{Nature}              % Nature
\def\araa{ARA\&A}             % Annual Review of Astronomy and Astrophysics
\def\pasj{PASJ}               % Publ.Ast.Soc.Japan
\def\prd{Phys. Rev. D}        % Phys.Rev.D
\def\prl{Phys. Rev. Lett.}    % Phys.Rev.D
\def\jcap{Journal of Cosmology and Astroparticle Physics}
\def\aapr{A\&A review}                % Astronomy and Astrophysics
\def\apss{Ap\&SS}                % Astronomy and Astrophysics

\section{Introduction}
\label{sec:intro}

GRB\,190114C is a long-duration GRB observed to emit gamma rays in the TeV band.
MAGIC detected, from about one minute after the burst, high-energy gamma rays ($\sim$0.2 to $\sim$0.8\,TeV) with high statistical significance, at the transition between the prompt and afterglow phases of the GRB emission \citep{Magicnature1}.
Several models have been proposed in order to reproduce the broadband emission observed in GRB\,190114C, both hadronic \citep{slf20, melandri22} and leptonic \citep{fraija19b,derishev19,ravasio19,wang19,chand20,rueda20}. Although, firm conclusions on the production mechanisms of GeV-TeV emission have not been reached so far, being also limited by the large number of parameters involved in GRB modelling. 

In this paper, we consider the hypothesis that part of the emission at high energy from GRB\,190114C could be caused by the presence of a hadronic component, with the subsequent production of high-energy neutrinos \citep{Eichler,Waxman,guetta15,yacobi14}. 

The GRB\,190114C light curve exhibits at the beginning irregular multi-peaks due to bunches of $\gamma$ rays superimposed on a smoothly varying emission component that extends beyond the highly variable emission period. Therefore observations do not exclude that part of the high energy emission may be released during the prompt phase.
We consider the possibility that a fraction of the  Very High-Energy (VHE) emission of GRB\,190114C may be due to photomeson interactions within the internal shocks (IS) region.
We consider the standard fireball model in which energy dissipation occurs at IS between shells in the relativistic outflow of the jet or through interactions with ambient matter \citep{Eichler,Waxman}. Thus, a substantial part of the bulk kinetic energy is converted into internal energy, which is then distributed between electrons, protons, and magnetic field. The internally accelerated electrons are presumably responsible for the keV–GeV photons observed in the GRB, which are emitted through synchrotron or inverse Compton processes. Accelerated protons may interact with these ($\sim$ MeV) photons and produce both neutral and charged pions, which in turn decay into high-energy photons and neutrinos, respectively. 
However, MAGIC observations did not
show any sign of significant spectral change or sharp flux variations from
prompt phase to prolonged emission that are characteristic features of the prompt phase. Furthermore, recent LHAASO observations \citep{lhaaso} provide a compelling evidence that the TeV emission can come from the afterglow, even if it overlaps in time with the
prompt emission. 
MAGIC observations started in the early afterglow phase, therefore several authors have interpreted this emission as due to the external shock model \cite{derishev19,slf20}.
In this paper we consider the possibility that part of the TeV emission is due to the internal shocks. 
The comparison between MAGIC data and the results of the Monte Carlo (MC) simulation developed for this work can set an upper limit on the contribution to very high energy photons due to the prompt phase.
The results described in this paper have been obtained by a full Monte Carlo simulation that 
described in detail in a previous paper \citep{fasano21}. For this simulation each accelerated proton is generated according to a power law ($E^{-2}$) and is tracked into a radiation field until it interacts or escapes. The radiation field is generated according to the Fermi measurements \citep{fermiGBM} as described in Sec.~\ref{sec:mc}. Photo-mesons interactions are simulated for the production of charged and neutral pions. Photons and neutrinos are so obtained from mesons decays. For each photon from pion decay we evaluate the probability to escape from the source, if the photon interacts we evaluate the probability that a secondary photon can emerge from the source. This makes our simulation different from previous works i.e.\citep{slf20}.
The paper is structured as follows: in Sec.~\ref{sec:grb}, we describe spectral and temporal properties of GRB\,190114C, assumed for the simulation. In Sec.~\ref{sec:mc}, we present our MC program, developed to simulate the photo-hadronic interactions occurring during the prompt phase. We also describe the result of the simulation of the electromagnetic cascades initiated by the interaction of high energy photons produced in $\pi^0$ decays. In Sec.~\ref{sec:results}, we compare the photon flux resulting from our simulation to MAGIC data, deconvolved for the Extragalactic Background Light (EBL) interactions. From this comparison we obtain a set of best fit values for our model. We then predict a flux of high energy neutrinos and evaluate the expected number of events in present and future experiments. A discussion of our results is provided in Sec.~\ref{sec:concl}.

\section{GRB\,190114C: spectral and temporal properties}
\label{sec:grb}
The first detection of GRB\,190114C is due to Swift \citep{swift}, further observations have followed by GBM \citep{fermiGBM} and LAT \citep{fermiLAT} onboard the \textit{Fermi} satellite (up to 22.9~GeV), AGILE/MCAL \citep{agile}, Integral/SPI-ACS \citep{integral} and Konus-Wind \citep{konus}. The combination of a rapid follow-up by MAGIC \citep{magicAtel} with the close distance of the source (confirmed from its optical counterpart to be at redshift $z=0.4245$ \citep{redshift1,redshift2}) allowed to unveil the presence of an extremely energetic radiation component in GRBs, as already expected by theory (e.g. \citep{TeVGRBs}).
The prompt phase of GRB\,190114C, as observed by \textit{Fermi}-GBM, appears as a multi-peak emission lasting $T_{90} \simeq116$~s (50-300~keV). The time-averaged spectrum in the first $\sim 40$~s can be described by a Band function with low and high-energy slopes equal to $\alpha=1.058$ and $\beta=3.18$ respectively, a break energy value $E_{\rm b} \simeq 1.1$~MeV in the observed frame \citep{fermiGBM}, and a gamma-ray fluence $F_\gamma=3.99 \times 10^{-4}$~erg~cm$^{-2}$ (10-1000 keV). As such, the isotropic energy release in the source frame amounts to $E_{\rm iso} \simeq 3 \times 10^{53}$~erg, indicating a fairly energetic GRB.
The MAGIC detection, with a significance above 50$\sigma$, occurred 68\,s after the \textit{Fermi}-GBM trigger. The VHE emission lasted for $\sim 40$ minutes, with the highest observed photon energy $E_{\rm max}=0.852$~TeV \citep{Magicnature1}. Within this temporal window, the time-dependent analysis of VHE data showed a systematic decrease in flux normalization, as well as a steepening trend over time. 
The high-energy photon spectrum reported by MAGIC, for the time interval 68-110~s, entirely overlaps with the $T_{90}$ estimated by \textit{Fermi}-GBM for the prompt emission. During this time interval, the intrinsic burst spectrum in the 0.2-1~TeV band is characterised by a pure power-law ($\propto E^{-\xi}$) with $\xi=2.16^{+0.29}_{-0.31}$~\citep{Magicnature2}. The photon spectral slope at the source has been derived deconvolving, from the observed spectrum the severe attenuation of the gamma ray flux due to its propagation to the Earth within the EBL \citep{Magicnature2}, according to the Dominguez et al. model \citep{dominguez}. We hence consider such an intrinsic spectrum and compare it to the prediction of the gamma-ray flux emerging from our simulation of phenomena happening in the IS region. Additionally, we investigated the effects of adopting a different EBL model, e.g. the one from Franceschini et al. \citep{franceschini1,franceschini2}, finding results consistent with the Dominguez et al. model within the statistical uncertainty of the MAGIC measurements. The temporal overlap between the MAGIC observations and $T_{90}$ is not sufficient to definitely attribute the high-energy photons measured by MAGIC to the prompt or to the afterglow phase of the GRB emission. 
In this paper we assume the hadronic scenario and we compare the MAGIC observation with the radiation emerging from photomeson interactions where accelerated protons interact with the Band-like target radiation field in the prompt phase. The two main features of the hadronic scenario are the bulk Lorentz factor $\Gamma$ of the relativistic jet, and the amount of energy channeled into relativistic protons $E_{\rm iso, p}$. This last quantity can be expressed with the baryon loading $f_{\rm p}=E_{\rm iso,p}/E_{\rm iso}$. Both the values of $\Gamma$ and $f_{\rm p}$ can be constrained to reproduce the VHE MAGIC observed spectrum.

\section{Monte Carlo simulation}
\label{sec:mc}

Modeling the physical processes occurring inside the IS region of the expanding GRB fireball requires characterization of the site where particles propagation and interactions take place.
Here we consider a simplified stationary one-zone scenario \citep{murase2006,asano2009,hummer2012} in which mildly relativistic shells of plasma collide at a typical radius \citep{Ris}
\begin{equation}
    R_{\rm IS}=\frac{2\Gamma^2 c t_{\mathrm{var}}}{(1+z)} \simeq 4 \times 10^{12} \, \left(\frac{\Gamma}{100} \right)^2 \, \left(\frac{t_{\rm var}}{0.01~{\rm s}} \right) \left(\frac{1.4}{1+z} \right) \, {\rm cm}
\end{equation}
where all the GRB energy is released.
As variability timescale, we assume the value $t_{\rm var}=6$~ms as suggested by observations during the prompt phase of GRB\,190114C \citep{ajello}. The bulk Lorentz factor $\Gamma$ is treated as a free parameter of the model, and it will be fixed as the one that best reproduces MAGIC data.\\
The MC calculation is performed in the IS frame, assuming a spherical geometry \citep{baerwald} and a shell width
$\Delta R_{\rm IS}= \Gamma c t_{\rm var} /(1+z) $.
We simulate a flux of accelerated protons with energies ranging from 1 GeV to $10^{9}$~GeV, 
according to $dN_{\rm p}/dE_{\rm p} \propto E_{\rm p}^{-2}$, consistently with a Fermi I-order acceleration process. Such a large energy range has been selected in order to avoid biases in the results.
The target photon energy distribution $dn_{\gamma}/d \epsilon_{\gamma}$ in the IS frame is assumed to reproduce the Band function observed by \textit{Fermi}-GBM \citep{fermiGBM} in the prompt phase. In order to allow the particle $\Delta^+$ resonance\footnote{Mass $m_\Delta=1232$~MeV/c$^2$, spin $S=0$, isospin $I=3/2$, total angular momentum $J=3/2$, parity $P=+$.} production with the entire spectrum of accelerated protons, the Band function has been extended to high energy following the trend of the higher part of the spectrum.
Each generated proton can either interact with ambient photons, if the center of mass energy is above the interaction threshold condition, or it propagates further in the shell. The average interaction length $\lambda_{p\gamma} (s) = [n_{\gamma} \sigma_{p \gamma}(s)]^{-1}$ is evaluated, depending on the $p\gamma$ center of mass energy and on the density of photons in the IS frame $n_\gamma(\epsilon_{\rm th,\Delta}^{\rm IS})$, requiring a center of mass energy above the $\Delta^+$ production threshold: $\epsilon_{\rm th,\Delta}^{\rm IS}=(m^2_\Delta-m^2_{\rm p})c^4/(4E_{\rm p}^{\rm IS})$, $m_\Delta$ and $m_{\rm p}$ being respectively the $\Delta^{+}$ and proton masses.

\begin{figure}[t!]
    \centering
    \includegraphics[scale=0.65]{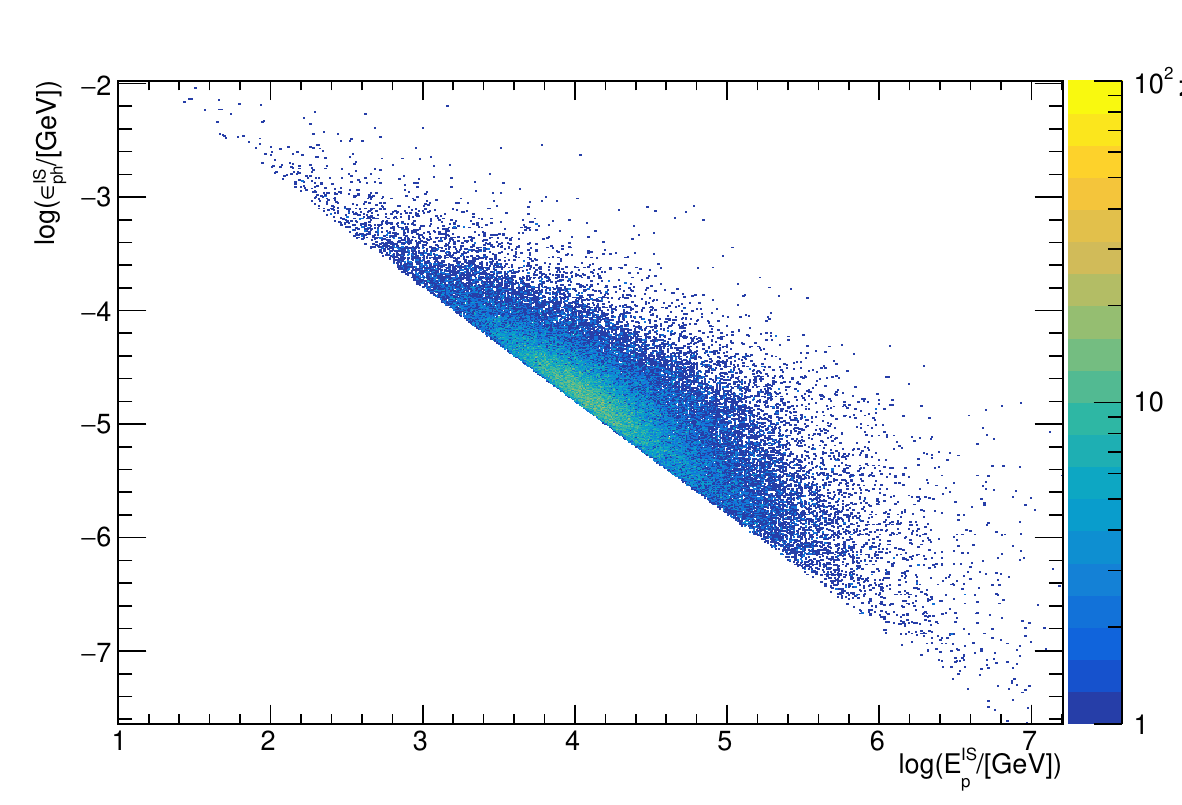}
    \caption{Target photon energy $\epsilon_{\rm ph}^{\rm IS}$ vs proton energy $E_{\rm p}^{\rm IS}$ for events above the $\Delta^{+}$ production threshold. Energies are given in the IS frame ($\Gamma=800$ and $t_{\rm var}=6$~ms). The colour code indicates the number of simulated events.}
    \label{fig:scattered}
\end{figure}

According to the average interaction lenght $\lambda_{p\gamma}(s)$ we extracted the proton path before the interaction $x_{\rm p}$. If $x_{\rm p} < R_{\rm IS}$ and if the threshold condition for the $\Delta^+$ production is satisfied, the photo-meson interaction occurs. The energies of protons ($E^{\rm IS}_{\rm p}$) and target photons ($\epsilon^{\rm IS}_{\rm ph}$) that satisfy the photo-meson production condition are shown in Fig.~\ref{fig:scattered}.

Multi-pion generation beyond resonant $\Delta^+$ production is also simulated and secondary particles like photons, muons, and neutrinos originated by decays are followed. %$\pi^0, \pi^{\pm}, \mu^\pm, \nu_{\mu}, \bar{\nu}_{\mu}, \nu_e, \bar{\nu}_e$, as well as secondary protons originated by the interaction, are followed by the simulation. 
Secondary protons, from $\Delta^+$ decay, are tracked until they leave the IS region to account for possible re-interactions along their path.

To account for proton energy losses during the propagation inside the IS region, we compare the proton acceleration timescale $t_{\rm acc}(E_{\rm p}^{\rm IS})=r_{\rm L}(E_{\rm p}^{\rm IS})/c$ ($r_{\rm L}$ being the particle Larmor radius), to  the average $\Delta^+$ production collision timescale $t_{p\gamma}(E_{\rm p}^{\rm IS})=[n_\gamma(\epsilon_{\rm th,\Delta}^{\rm IS}) c \sigma^{\Delta}_{p\gamma} K_{p\gamma}]^{-1}$. In the previous expression, $K_{p\gamma}=0.13$ and $\sigma^{\Delta}_{p\gamma} = 5\times 10^{-28}$~cm$^2$ are respectively the inelasticity coefficient and the cross section for the interaction at the threshold \citep{atoyan}.

Charged particles may be affected by synchrotron losses,
therefore, in order to evaluate these effects, we 
calculate the magnetic field value at IS by equating the magnetic energy density $U_{\rm B}=B^2/(8\pi)$ to the kinetic energy density of the accelerated electrons. 
Naming $\epsilon_{\rm B}$ and $\epsilon_{\rm e}$ respectively the fraction of jet energy converted into magnetic field and carried by electrons, we assume $\epsilon_{\rm e}=\epsilon_{\rm B}=0.1$ \citep{waxman_epsilon}.
Proton synchrotron losses are relevant for energy above $E_{\rm p, cut}^{\rm IS}\simeq 45, 65, 100, 400$~PeV assuming $\Gamma=300, 500, 800, 1000$ and $t_{\rm var}=6$~ms. These evaluations confirm our choice to simulate the proton energy up to $10^9$ GeV. Pion and muon propagations into magnetic field are also affected by synchrotron energy losses. We have compared the particles lifetime with the synchrotron timescale, $t_{\rm syn}(E^{\rm IS})=3m_x^4 c^3 /4 \sigma_T m_e^2 \epsilon_{\pi} U_B$, \citep{guettagranot}, where $m_x$ is the mass of the particle which we are taking into account. We found $E^{\rm IS}_{\mu^\pm, \rm cut} \simeq 0.5,2.6,10,20$~TeV and $E^{\rm IS}_{\pi^\pm, \rm cut} \simeq 10,50,200,400$~TeV ($\Gamma=300,500,800,1000$, $t_{\rm var}=6$~ms).

\begin{figure}[t]
    \centering
    \includegraphics[scale=0.65]{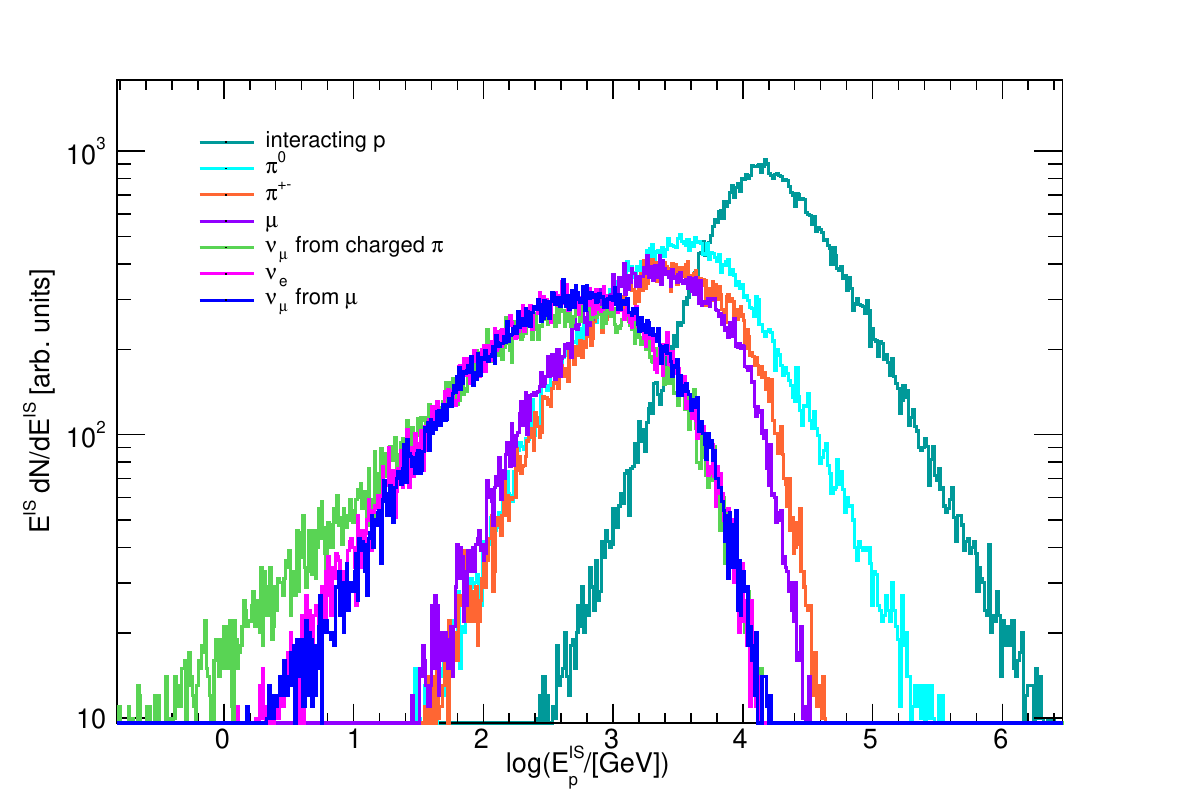}
    \caption{Particle spectra resulting from p$\gamma$ interactions, obtained in the IS frame from the simulation with $\Gamma=800$ and $t_{\rm var}=6$~ms.}
    \label{fig:spectrum}
\end{figure}
\noindent
The MC simulation of $p\gamma$ interactions follows the methods described in \citep{fasano21}. The energy particle distribution for interacting protons, charged pions, muons and neutrinos assuming $\Gamma=800$ and $t_{\rm var}=6$~ms is shown in Fig.~\ref{fig:spectrum}.

Concerning the neutral pion production and decay, we consider the possibility that each originated gamma ray might escape from the IS region, or interact with target photons, via $e^\pm$ pair production. The average radiation length for this process is evaluated following \citep{art:waxman2003} as
\begin{equation}
\lambda_{\gamma \gamma}^{-1} (E_{\gamma}^{\rm IS}) = \frac{1}{2} \sigma_{\gamma \gamma} \int d \cos\theta (1-\cos\theta) \int_{\epsilon_{th}}^{\infty} U_{\gamma}(\epsilon_{\gamma}) \frac{1}{2} \frac{dn_{\gamma}}{d\epsilon_{\gamma}} d\epsilon_{\gamma}    
\label{eq:lambda_waxman}
\end{equation}

\noindent 
with the condition $\epsilon_{th} E_{\gamma}^{\rm IS} (1-\cos \theta) > 2(m_e c^2)^2$, being $0<\theta<\pi$ the angle between the photons in the center of mass reference frame, and $\sigma_{\gamma \gamma}$ the $e^\pm$ pair production cross section \citep{vernetto}. For the target photons we assumed the spectral energy distribution $dn_{\gamma}/d\epsilon_{\gamma}$ approximated by a Band function as described in Sec. \ref{sec:grb}. For each $\pi^0$-originated photon with energy $E_\gamma^{\rm IS}$, we extract its free path $x_{\gamma}$ according to the average radiation length $\lambda_{\gamma \gamma}$ in Eq.~\eqref{eq:lambda_waxman}. If $x_\gamma> R_{\rm IS}$ the photon escapes the IS region and will eventually be observed in the laboratory frame with energy $E_{\rm obs}= \Gamma E_{\rm IS}/(1+z)$. Most of the photons originated by $\pi^0$ decays are absorbed and initiate an electromagnetic cascade. We track the products of the interactions, until their energy can be
included within the MAGIC observed energy range.

The contribution of photons from electromagnetic cascades to the MAGIC observed spectrum depends on the bulk Lorentz factor value used in the simulation: for a given $\pi^0$ energy this contribution from cascades is larger for lower values of $\Gamma$.
The flux of escaping photons can be directly compared with the intrinsic spectral energy density evaluated on the basis of  MAGIC published results \citep{Magicnature2}. The contribution of photons from electromagnetic cascades has been considered 
previously for GRBs \citep{wang} and other astrophysical sources (i.e. blazars \citep{Bottcher}). For our work, instead of the semi-analytical treatment described in \citep{Kelner}, we have developed a full MC simulation and applied it to GRB190114C.

\section{Results: gamma rays and neutrinos from photo-hadronic interactions}
\label{sec:results}
We perform MC simulation for different values of $t_{\rm var}$ and $\Gamma$, obtaining spectra of high-energy gamma rays and neutrinos emerging from the interaction region.  
We then convert these spectra into expected fluxes on Earth by taking into account the cosmological nature of GRBs. 
We evaluate the quantity $E_\gamma^2\phi(E_\gamma)$ on Earth rescaling the IS energy spectrum according to:
i) %the comoving distance $d_{\rm c}=d_{\rm L}/(1+z)$, where
the luminosity distance $d_{\rm L}$,
ii) the first time interval of MAGIC observations $\Delta t=42$~s, and iii) the energy conversion factor from the IS to the observed frame, as: 
\begin{equation}
E_\gamma^2 \phi(E_\gamma)= f_{\rm p} \Gamma \frac{(1+z)}{4\pi d_{\rm L}^2 \Delta t} \left( E_\gamma^2 \frac{dN_\gamma}{dE_\gamma} \right)_{\rm IS} \, .
\label{magic:4}
\end{equation}

For each simulated $\Gamma$ and $t_{\rm var}$, the baryon loading value $f_{\rm p}$ has been assumed in the interval $1-30$ %in uniform steps of width $\Delta f_{\rm p}=1$,
to find the best agreement between the EBL-deconvolved MAGIC data, in the 68-110~s time interval \citep{Magicnature2} and the simulated fluxes, through a $\chi^2$ statistical test. For the assumed values of $\Gamma=300,500,800,1000$ we obtain $f_{\rm p}=23 \pm 6, 13 \pm 3 ,11 \pm 2, 7 \pm 2$ as best-fit parameters. These values are well consistent with the expectations \citep{murase_baryon}.

\begin{figure}[t]
    \centering
    \includegraphics[scale=0.65]{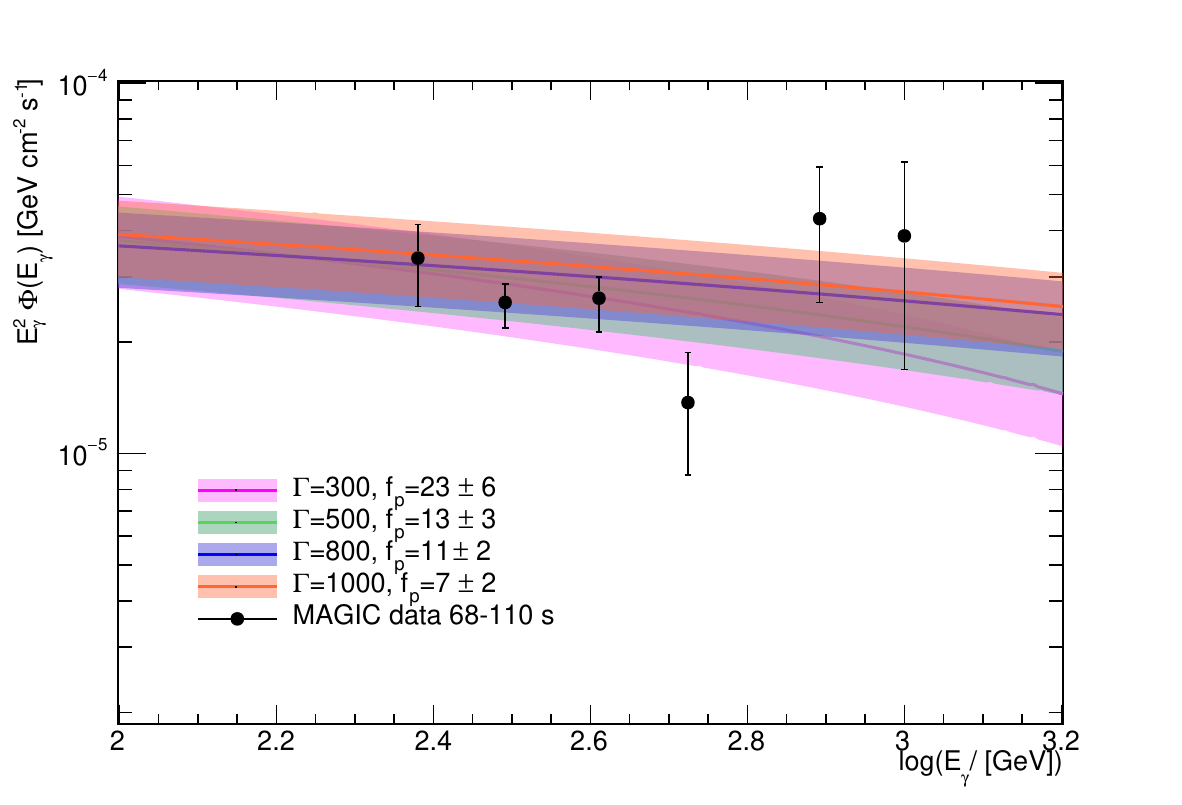}
    \caption{IS simulation: comparison between the EBL-deconvolved flux of GRB\,190114C, as provided by MAGIC in the temporal interval 68-110~s \citep{Magicnature2} and the MC results. Photon fluxes are due to $\pi^0$-decays and to electromagnetic showers following internal gamma-ray absorption on the IS for different parameters of the model.}
    \label{fig:magic_vs_simulation}
    
\end{figure}

For each $\Gamma$ value, we evaluated the expected flux on Earth $E_{\gamma}^2 \Phi(E_\gamma)$. This result is shown in Fig. \ref{fig:magic_vs_simulation} where the shaded area represents the $1\sigma$ statistical uncertainty on $f_{\rm p}$. For the assumed values of $\Gamma$, we found that the majority of photons emerging from the source is originated in the electromagnetic cascades due to interactions of photons from $\pi^0$.\\
A direct proof of the hadronic origin of the observed TeV radiation might come from coincident neutrino observations. Because of the transient nature of GRBs, the detection of a single neutrino event would allow identifying these sources as extreme hadronic accelerators. 

\begin{figure}[t]
    \centering
    \includegraphics[scale=0.65]{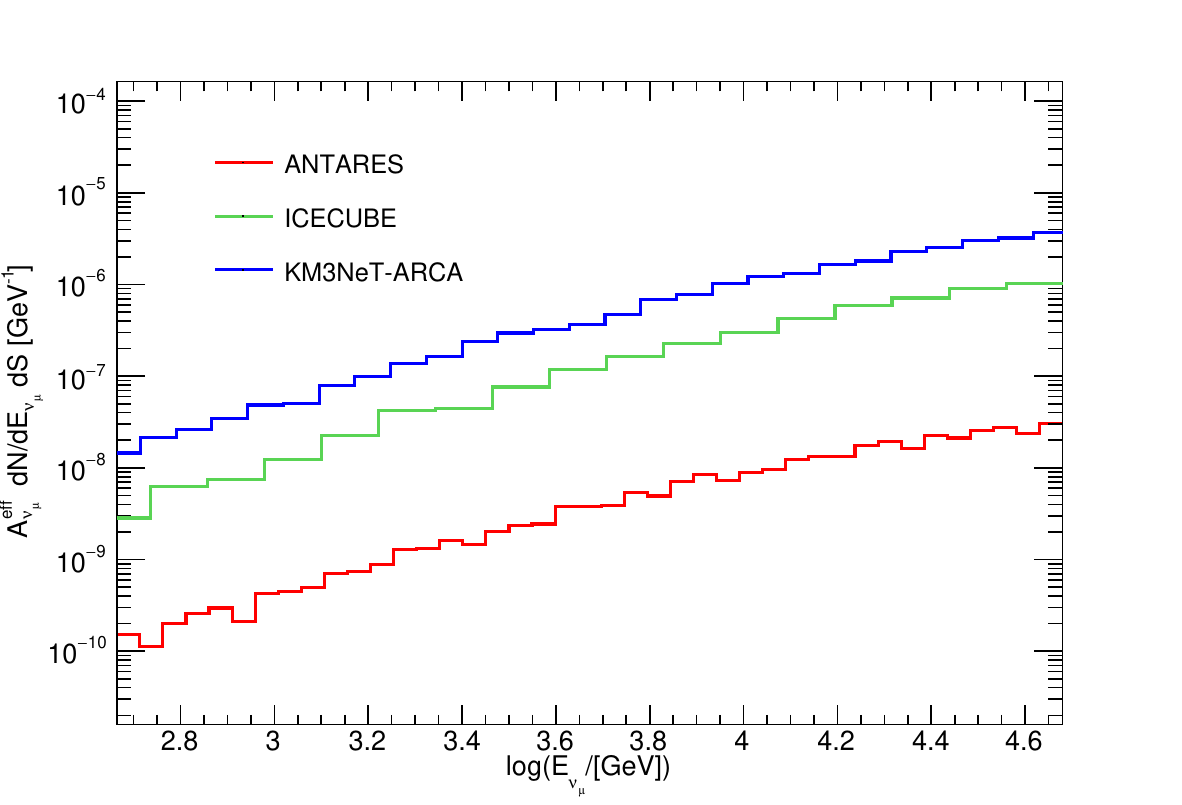}
    
    \caption{Number of expected  muon neutrino and antineutrino events, for each energy bin, 
    from GRB190114C for $\Gamma=800$ and $t_{\rm var}=6$ ms, and $f_{\rm p}= 11$: red for ANTARES \citep{aeffA}, green for Icecube \citep{aeffI}, blue for KM3NeT/ARCA \citep{loi}.}
     \label{fig:neutrini}
    
\end{figure}

To evaluate the neutrino energy flux on Earth for muon neutrinos and antineutrinos we assume $\Gamma=800$, we use the energy conversion factor described in Eq.\eqref{magic:4}, and we
take into account neutrino oscillations, assuming normal ordering and the standard three-flavor scenario \citep{vissani}.
To check whether present and future neutrino detectors might be able to find a signal from an astrophysical source similar to a GRB\,190114C, we have evaluated the expected neutrino, plus antineutrino, fluence $dN_{\nu_\mu+\bar{\nu_\mu}}/dE_{\nu_\mu+\bar{\nu_\mu}} dS$. We can then compute the expected number of track-like events with the 
\begin{equation}
N_{\rm events}(\delta)= \int A_{\rm eff}^{\nu_\mu} (E_{\nu},\delta) \left( \frac{dN_{\nu_\mu}}{dE_{\nu_\mu}dS} \right)_{\rm Earth} dE_{\nu_\mu} \, ,
\end{equation}
where $A_{\rm eff}^{\nu_\mu} (E_{\nu_{\mu}}, \delta)$ is the Neutrino Telescope effective area, given as a function of the neutrino energy and of the source declination $\delta$.
Public effective areas of ANTARES \citep{aeffA} and IceCube \citep{aeffI} are evaluated at analysis level and for the declination band that includes the position of GRB\,190114C. For KM3NeT/ARCA the effective area is evaluated at trigger level and averaged throughout the sky  \citep{loi}.
In Fig.~\ref{fig:neutrini} we report the number of muon neutrino and antineutrino expected events for the three neutrino detectors as a function of the neutrino energy. In Table \ref{tab:eventi_attesi}, we provide the total number of neutrino-induced events expected in the three detectors, obtained by the simulation with $t_{\rm var}=6$~ms, $\Gamma=800$ and $f_{\rm p}=11$. 
Both the ANTARES and IceCube Collaborations have unsuccessfully searched for coincident neutrino-induced signals from the direction of GRB\,190114C.  In the case of ANTARES, the derived 90\% confidence level integrated upper limit amounts to $1.6$~GeV/cm$^2$.
For IceCube the same limit amount to $0.44$~GeV/cm$^2$ \citep{antares2021, ICatel}. In both cases, the constraints are limited to the muon neutrino component reaching Earth. In fact, because angular precision is a crucial feature in reducing the atmospheric background entering the search cone angle, muon neutrino charged-current interactions constitute the better astronomical channel, as muons emerging from these can be identified in neutrino telescopes as long tracks. 
The non-detection of current instruments is compatible with the hadronic model expectations for the values of $f_{\rm p}$ and $\Gamma$ that better reproduce the sub-TeV gamma-ray MAGIC data \citep{icecube_followup}.

\begin{table}[h]
\begin{center}
\begin{tabular}{ccc}
\hline
Detector & Declination band & $N_{\rm events}$ \\
\hline
ANTARES & $-45^o<\delta<0^o$ & $1 \times 10^{-3}$ \\
IceCube & $-30^o<\delta<0^o$ & $3 \times 10^{-2}$\\
KM3NeT/ARCA & Average & $1 \times 10^{-1}$\\
\hline
\end{tabular}
\caption{Number of neutrino induced events expected by the GRB Internal Shock MC simulation for different neutrino telescopes, as due to $\nu_\mu$ and $\bar{\nu}_\mu$ interactions during the [68;100]~s time interval of the prompt emission of GRB\,190114C, for the model with $t_{\rm var}=6$~ms, $\Gamma=800$ and $f_{\rm p}=11$.}
\label{tab:eventi_attesi}
  
\end{center}
\end{table}

Since the coincident background rate expected from atmospheric neutrinos is by orders of magnitude lower than the signal rate expected from GRB\,190114C, a detection from a single GRB appears to be very difficult. This result suggests that stacking several GRBs of the same kind could be a viable solution for a significant detection.
Despite the experimental efforts, no clear signal has emerged so far in data from the ANTARES and IceCube telescopes \citep{aeffA,aeffI} when searching for spatial and temporal coincidences with the prompt emission of a sample of stacked GRBs. This leads to a constraint on the possible contribution of standard GRBs to less than 10\% of the diffuse cosmic neutrino flux \citep{antGRB1,antGRB2,antGRB3,icGRB}.
Furthermore, it is still unclear whether TeV-emitting GRBs behave as the standard GRB population.

\section{Discussion and Conclusions}
\label{sec:concl}
The successful observation of radiation from GRBs extending up to the TeV domain has provided further evidence of the extreme nature of these sources. 
However, there is still poor understanding of the processes that characterize these TeV emissions that might witness the presence of effective hadronic acceleration in GRB jets, possibly already during the prompt phase of the emission or, more realistically, during the afterglow phase. This occurrence would establish the connection between GRBs and Ultra-High-Energy Cosmic Rays (UHECRs), which remains a long-standing paradigm still to be proven.

To test the hypothesis of the hadronic origin of TeV radiation, we developed a MC simulation of photo-meson interactions between high-energy protons and target photons distributed according to a Band-like spectrum, as indicated by Fermi-GBM observations, during the prompt phase. 
After computing the spectra of secondary particles emerging from these interactions, we additionally simulated the electromagnetic absorption that gamma rays undergo in the IS shell. The spectrum of escaping photons so obtained has been compared to the intrinsic source spectrum derived by the MAGIC observations of GRB\,190114C.
We obtained our results by developing a full MC simulation of the interactions of high energy accelerated protons with the IS radiation. High energy photons from $\pi^0$, are followed until they interact in the IS region or escape. The total intrinsic flux of photons, directly from $\pi^0$ or from electromagnetic cascades, escaping from the interacting regions is then evaluated. This flux transported to Earth is compared with MAGIC observations deconvolved by EBL interaction. Several simulations have been performed, in the framework of both IS scenarios, assuming different values of the relevant model parameters, like the Lorentz factor $\Gamma$ and the baryon loading $f_p$.
From the comparison of the simulation results and the MAGIC data, we extracted the parameter values that better reproduce the observations. 
In this paper we have considered the possibility that part of the TeV emission may be due to the IS model responsible of the prompt phase. We have found the best fit parameters for this model that can reproduce the maximum contribution due to the IS interactions.

The same MC code has been used to evaluate the flux of neutrino from charged pion decay arriving at Earth. Applying the same MC simulation to the entire sample of the observed TeV GRBs could provide insights on the physical mechanisms responsible for the TeV emission.
A better knowledge of the high energy photon spectrum, that at present seems to be limited by the late response of imaging atmospheric Cherenkov telescopes in pointing, would help in the GRB characterization. 
Confirmation of the hadronic origin of sub-TeV radiation is expected from neutrino observations. 
However our simulation indicates that such a detection from individual astrophysical sources like GRB\,190114C appears extremely unrealistic, as confirmed by the lack of spatial correlations in data reported by the ANTARES and IceCube neutrino telescopes.

\begin{acknowledgments}
The authors acknowledge the support of the Amaldi Research Center funded by the MIUR programme  “Dipartimento  di  Eccellenza”  (CUP:B81I18001170001),  the  Sapienza  School  for Advanced Studies (SSAS) and the support of the Sapienza Grants No. RM120172AEF49A82 and RG12117A87956C66. The authors are thankful to Enrico Peretti for fruitful discussions regarding the manuscript content.
\end{acknowledgments}

{}

\end{document}